\documentclass[aps,prl,preprint,groupedaddress]{revtex4}
\usepackage{graphicx}
\begin{document}

%\preprint{}

\title{Resistance between two nodes in general position on an $m \times n$ fan network }


\author{ J. W. Essam  \footnote{E-mail: j.essam@rhul.ac.uk }}
\address{  Department of Mathematics, Royal Holloway College, University of London, Egham, Surrey TW20 0EX, England.}

\author{ Zhi-Zhong Tan \footnote{E-mail: tanz@ntu.edu.cn ; ~~ tanz@163.com}}
\address{  Department of Physics, Nantong University, Nantong 226007, China.}

\author{F. Y. Wu \footnote{E-mail: fywu@neu.edu }}
\address{  Department of Physics, Northeastern University, Boston, MA 02115, USA.}

\date{\today}

\begin{abstract}
The resistance between two nodes in general position on a fan network with $n$ radial lines and $m$ transverse lines is determined.  Also a similar result of Izmailian, Kenna and Wu [7] for an $m\times n$ cobweb network is reproduced but the method used here is significantly different. It avoids the use of the Kirchhoff matrix, requires the solution of just one instead of two eigenvalue problems and results directly in only a single summation. Further the current distribution is given explicitly as a biproduct of the method. The method is the same as that  used by Tan, Zhou and Yang [10] to find the cobweb resistance between center and perimeter for $1\le m\le3$ and general $n$.  Proof of their conjecture for general $m$ is discussed.


\pagenumbering{arabic}

\end{abstract}

\pacs{ }

\maketitle


\section{1. Introduction}
Recently there has been increasing interest in the point to point resistance of finite resistor networks. For an up to date list of references see [1]. Wu [2] expressed the resistance of a finite graph in terms of the eigenvalues and eigenvectors of the Kirchhoff matrix [3] avoiding the zero eigenvalue. He also used his method to obtain the resistance of a rectangular grid with various boundary conditions. Recently Chair [4],[5] has used Wu's formula to obtain exact formulae for the complete graph minus N edges and the N-cycle graph with first and second neighbours. Essam and Wu [6] used one of the results of [2] to find an asymptotic expansion for the resistance between opposite corners of a rectangular network with free boundary conditions. This was extended by Izmailian and Huang [1] to other boundary conditions. Izmailian, Kenna and Wu [7] have recently introduced a modification of a the method of Wu [2]. Their new method also avoids the zero eigenvalue of the Kirchhoff matrix by expressing the resistance in terms of the eigenvalues and eigenvectors of a principal cofactor. This method will be referred to here as the IKW method.

Current interest lies in rectangular networks with zero resistance along one edge. Here the case of free boundary conditions along the other three edges will be considered. Contracting the zero resistance branches leads to a network which can be displayed as a fan (see figure 1). The $m\times n$ fan will be supposed to have n radial lines and m transverse arcs. The main topic of this article is the derivation of the resistance between two nodes of a fan network in general position. The result is shown in equation (2). The special case with one node at the apex and the other on the boundary has appeared in [8].

\begin{figure*}
\begin{center}
\includegraphics[width=9cm,bb=0 0 323 190]{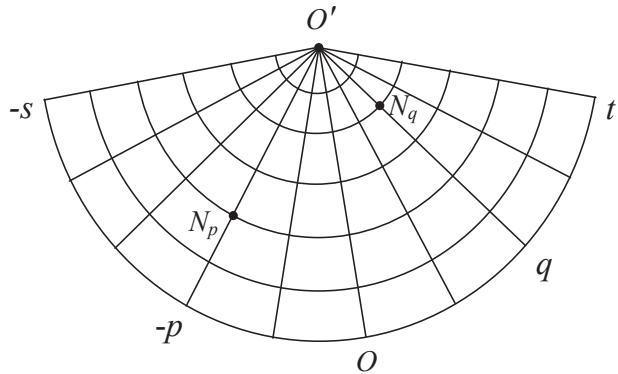}
\caption{ \it The $m=6, n=8, y_{p}=2, y_{q}=4$ fan network. }
\end{center}
\end{figure*}

Tan and coworkers [9], [10], [11] introduced a cobweb network which has periodic boundary conditions in the direction parallel to the zero resistance edge. The case $m = 6, n = 8$ of an $m\times n$  cobweb network is shown in figure 2. The cobweb network is related to the fan network by joining corresponding nodes on the first and last radial lines of the fan. The formula for the symmetric case of the fan with n odd and both input and output nodes on the central radial line yields the result for the cobweb with both nodes on any single radial line. The cobweb network may also be thought of as a network embedded on the surface of a cylinder with resistors along the length and round the circumference and one end closed.

Izmailian, Kenna and Wu [7] used their Kirchhoff matrix method to obtain the resistance of the cobweb network between two nodes in general position. The IKW method requires the solution of two eigenvalue problems and results in a double summation. For the fan resistance considered here we use the method of Tan, Zhou and Yang [10] which requires the solution of just one eigenvalue problem together with the solution of a linear recurrence relation with constant coefficients. The final formula (2.2) involves only a single summation. By a simple change of boundary condition the cobweb formula of [7] is reproduced.

Tan, Zhou and Yang [10] considered the resistance of an $m\times n$ cobweb network between a node at the center and a node on the boundary. On the basis of results for general n and m = 1, m = 2 [9] and m = 3 [10] they conjectured the result for general m and n (see section 2.2.1). The case m = 4 was proved by Tan, Zhou and Lou [11]. The conjecture for general m and n has recently been proved by Izmailian, Kenna and Wu [7] as a special case of their formula for the resistance between an arbitrary pair of nodes on the cobweb. Further analysis was required to reduce their double summation to the single summation formula of the conjecture.

A proof of the above conjecture for odd n by the present authors [8] results from restricting the calculation to the resistance between the apex and a general node on the boundary of the fan. As noted above the resistance when the node of the fan is in the midpoint of the boundary is the same as that of the cobweb. The method used was a simplification of the one used here. The proof [8] was shorter than that of Izmailian et al [7] due to the simpler method and restricted domain.
\begin{figure*}
\begin{center}
\includegraphics[width=8cm,bb=0 0 416 351]{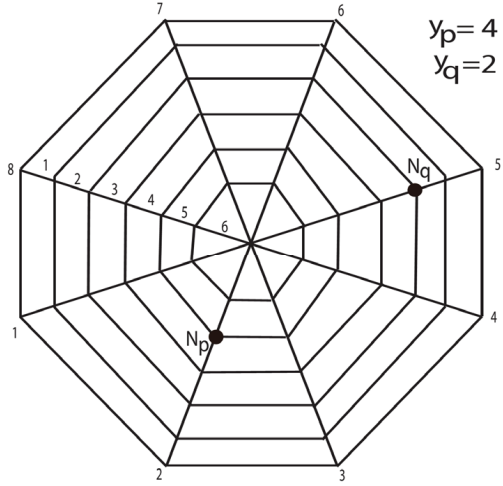}
\caption{ \it A $6\times 8$ cobweb network }
\end{center}
\end{figure*}
\subsection{1.1 The layout }
Formulae for both the fan (2) and cobweb network (6) with input and output nodes in general position are presented in section 2. Also simplifications of the formulae are considered when both nodes are on the same radial or transverse line. In section 3 the details of the method are given in the context of the fan. Section 4 gives the change in boundary condition required to obtain the cobweb formula. The results are discussed in section 5 where the method used here [10] is summarised and compared with the IKW method [7].
\section{2. Results }
\subsection{2.1 Resistance of the $m\times $ n fan and cobweb networks between two nodes in general position}
In general the fan has $n=s+t+1$ radial lines labelled from $k=-s$ to $k=t$.  Its resistance $R_{fan}$ will be determined between two nodes, one, $N_q$, distant $y_q$ up the radial line $k=q$ and the other, $N_p$, distant $y_p$ up the radial line $k=-p$ (see figure 1).

Let $\theta_{i}=\frac{2i-1}{2k+1}\frac{\pi}{2}$, $v_{i}=2\cos(2\theta_{i}) $ and $u_{i}=2+\frac{r}{r_{0}}(2-v_{i})$. Further suppose that $\lambda_{i}, \bar{\lambda}_{i}$ are the greater and lesser solutions of
\begin{eqnarray}
     \lambda_{i}^{2}-u\lambda_{i}+1=0 .
 \end{eqnarray}
With $L_i =\frac 12\log \lambda_i$ the result is
\begin{equation}
R_{fan} = \frac{r}{2m+1}\sum_{i=1}^m\frac {\alpha_{fan} C_{i}(y_{q})^2 -2\beta_{fan} C_{i}(y_{q})C_i(y_{p}) +\gamma_{fan} C_i(y_{p})^2}{\sinh(2L_{i})\sinh(2nL_{i})} ,
\end{equation}
where  $C_{i}(y) = \cos(2y+1)\theta_{i}$ and
\begin{eqnarray}
\alpha_{fan}  =4\cosh(2t-2q+1)L_i \cosh(2s+2q+1)L_{i} ,\\
\beta_{fan}  =4\cosh(2t-2q+1)L_{i} \cosh(2s-2p+1)L_{i} ,\\
\gamma_{fan}  =4\cosh(2t+2p+1)L_{i} \cosh(2s-2p+1)L_{i} .
\end{eqnarray}
For the cobweb network the $k=-s$ and $k=t$ lines are joined by additional branches to create cylindrical boundary conditions. The result is independent of $s$ and $t$ and depends only on $y_p$, $y_q$ and the transverse distance $d = p+q$ between $N_p$ and $N_q$ as expected. In section 4 the fan proof is simply extended to the cobweb by changing only the boundary condition with the result
\begin{eqnarray}
R_{cob }= \frac{r}{2m+1}\sum_{i=1}^m \frac {\alpha_{cob}( C_i(y_q)^2+C_i(y_p)^2) -2\beta_{cob} C_i(y_q)C_i(y_p)}
{\sinh(2L_i)\sinh(nL_i)} ,\\
\alpha_{cob}=2\cosh(nL_i)\qquad\hbox{and} \qquad \beta_{cob} = 2\cosh[(n-2d)L_i].
 \end{eqnarray}
Equation (6) is equivalent to equation (30) of Izmailian, Kenna and Wu [7]. Their summation runs from $i=0$ to $i=m-1$ and in their notation $L_i=\Lambda_{i-1}$, $\theta_i =\phi_{i-1}$ and $y\rightarrow m-y$.

\begin{equation}
\cosh(2L_i) = \frac {\lambda_i+\bar \lambda_i}2 = \frac 12 u_i =1 + \frac {2r}{r_0} \sin^2 \theta_i ,
\end{equation}
so that $L_i$ and hence $R_{fan}$ and $R_{cob}$ are determined by the resistance ratio and $\theta_i$ apart from the parameters $m,n,s,t,p,q$.

The formulae may be simplified in the following two special cases which for the cobweb already appear in [7]
but with different notation.

\subsection{2.2 Resistance between two nodes on the same radial line}
Without loss of generality we take the line to be $k=0$ and set $p=q=0$. In this case $\alpha_{fan} = \beta_{fan} = \gamma_{fan}$ with the result
\begin{equation}
R_{fan}^{radial}=\frac{4r}{2m+1} \sum_{i=1}^m [C_i(y_q)-C_i(y_p)]^2 \frac{\cosh[(2s+1)L_i]\cosh[(2t+1)L_i]}{\sinh[2L_i]\sinh[(2n L_i]} .
\end{equation}
Note: $p = q = 0$ means that $N_{p}$ and $N_{q}$ are both on the $k = 0$ radial line but should not be taken to mean $y_{p} = y_{q}$ as the notation suggests. If $N_{p}$ is the apex then $y_{p} = m$ and $C_{i}(y_{p}) = 0$.
Setting $d=0$ in (6) also gives the the resistance between two nodes on the same radial line for the cobweb.
\begin{equation}
R_{cob}^{radial}=\frac{2r}{2m+1} \sum_{i=1}^m [C_i(y_q)-C_i(y_p))]^2 \frac{\coth[nL_i]}{\sinh[2L_i]} ,
\end{equation}
which is equal to $R_{fan}^{radial}|_{s=t}$.  This is because in this symmetric case of the fan the potentials on the left and right radial boundaries are equal.  Thus when the nodes on the left radial line are joined to the corresponding nodes on the right radial line no current flows through the connecting branches. The resistance of the cobweb network so formed is therefore the same as for the fan.

\subsubsection{2.2.1 Proof of the Tan, Zhou and Yang conjecture for the cobweb network}
 Tan et al [10] conjectured that the resistance from center to the boundary of the cobweb network is given by
 \begin{equation}
R_{cob}=\frac{r}{2m+1} \sum_{i=1}^m(2+v_{i})\frac{\coth n\ln\sqrt{\lambda_{i}}}{\lambda_{i}-\bar{\lambda}_{i}} ,
\end{equation}
where $v_{i}=2cos[\frac{2i-1}{2m+1}]$. Since $y_{p}=0 $ and $y_{q}=m $, $[C_{i}(y_{q})-C_{i}(y_{p})]^{2}=(\cos\theta_{i})^{2}$. Thus from (10)
\begin{equation}
R_{cob}^{radial}|_{y_q=0,y_p=m}=\frac{2r}{2m+1} \sum_{i=1}^m (\cos \theta_i)^2 \frac{\coth[nL_i]}{\sinh[2L_i]} .
\end{equation}
But $2+v_i = 2(1+\cos 2\theta_i)$ and $\sinh(2L_i) = 2 (\lambda_{I} -\bar \lambda_{I})$ which proves the conjecture. This is the proof of Izmailian, Kenna and Wu [7].

\begin{figure*}
\begin{center}
\includegraphics[width=9cm,bb=0 0 161 89]{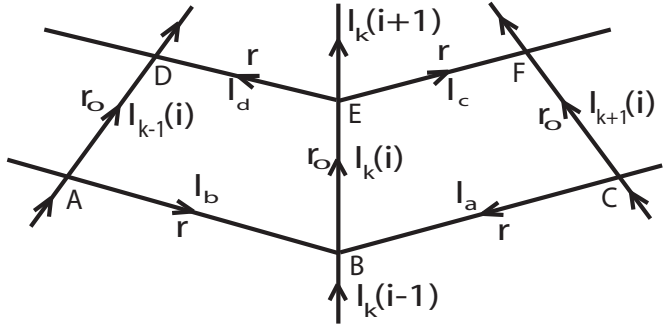}
\caption{ \it The voltage loop ABEFCBEDA.}
\end{center}
\end{figure*}

\subsection{2.3 The case of input and output nodes at the same distance from the apex}
Setting $y_p=y_q=y$ implies $C_i(y_p) = C_i(y_q)$ so that the numerator of the summand in (2) becomes $(\alpha_{fan}-2\beta_{fan}+\gamma_{fan})C_i(y)^2$. Further if we set $p=q$ the distance between the input and output nodes is $d=2q$.
In this case
\begin{equation}
\alpha_{fan}-2\beta_{fan}+\gamma_{fan} = 2\sinh(dL_i)(\sinh [(2n-d)L_i] +\sinh(dL_i) \cosh [2(t-s)L_i]).
\end{equation}
If further $s=t$ then the input and output nodes are symmetrically placed relative to the radial line boundaries and
\begin{equation}
R_{fan}^{trans}|_{s=t}= \frac{8r}{2m+1}\sum_{i=1}^m\frac { \sinh(d L_i)\cosh[(n-d)L_i]}{\sinh(2L_i)\cosh(nL_i)} C_i(y)^2 .
\end{equation}
Setting $y_p=y_q=y$ in (3)
\begin{equation}
R_{cob}^{trans}= \frac{8r}{2m+1}\sum_{i=1}^m\frac { \sinh(d L_i)\sinh[(n-d)L_i]}{\sinh(2L_i)\sinh(nL_i)} C_i(y)^2 .
\end{equation}
In this case $R_{cob}^{trans}\ne R_{fan}^{trans}|_{s=t}$ since the symmetry is destroyed by the net current flow from source to sink  in the transverse direction.

\section{3. Derivation of the fan formula (2)}
Firstly the formula for the fan will be derived and that for the cobweb will follow immediately by a simple change of boundary condition.

To find $R_{fan}$ we inject current $J$ at node $N_q$ and remove it  at $N_p$. Let $I_k(i)$ be the resulting current in the $i^{th}$ resistor from the edge of the $k^{th}$ radial line flowing towards the center (see figure 3). Using Ohm's law  the potential difference may be measured along a path from $N_q$ to $0$ and then to $N_p$ with the result
\begin{equation}
R_{fan}= \frac{r_0} {J}\left( \sum _{i=y_q+1}^m I_q(i)- \sum _{i=y_p+1}^m I_{-p}(i)\right ) .
\end{equation}

\subsection{3.1 Relating the current distribution in three adjacent radial lines}
To determine the radial currents consider the voltage loop $ABEFCBEDA$, shown in figure 1, centered on the $i^{th}$ resistor of the $k^{th}$ radial line .  If current $J$ enters at the node of height $y$ on the radial line $k=x$ charge conservation gives

\begin{eqnarray}
I_a+I_b &=I_k(i) -I_k(i-1) -J \delta_{i,y+1}\delta_{k,x}  \\
\hbox{and}\qquad  I_c+I_d&= I_k(i)-I_k(i+1) +J\delta_{i,y}\delta_{k,x} .
\end{eqnarray}
When $i=1$ in (17), $I_k(i-1) = 0$.
The sum of the voltage differences round the loop is zero so using Ohm's law
\begin{equation}
r_0(2I_k(i) - I_{k-1}(i) -I_{k+1}(i)) +r(I_a+I_b) +r_1(I_c+I_d)=0 ,
\end{equation}
where $r_1=r$ for $i<m$ and is zero for $i=m$. Combining these equations
\begin{eqnarray}
I_{k+1}(i) =&-h I_k(i-1)+ (h+h_1+2)I_k(i)   -h_1 I_k(i+1) -I_{k-1}(i)\nonumber\\
 &+J(h_1\delta_{i,y}-h\delta_{i,y+1})\delta_{k,x} ,
\end{eqnarray}
where $h=r/r_0$, $h_1= r_1/r_0$.

Equation (20) may be written in matrix form
\begin{equation}
	I_{k+1}= [(2h+2)U_m - h V_m]I_k - I_{k-1} -hJ\delta_{k,x}\epsilon_{i,y} ,
\end{equation}
where $U_m$ is an $m$-dimensional unit matrix, $\epsilon_y$ is a column matrix with $i^{th}$ element $\epsilon_{i,y}= \delta_{i,y+1} -\delta_{i,y}$ and
\begin{equation}
{\tiny V_m = \left (\begin{array}{lllllllll}
0&1&0&0&\dots&0&0&0\\
1&0&1&0&\dots &0&0&0\\
0&1&0&1&\dots &0&0&0\\
\vdots&\vdots&\vdots &\vdots&\dots&\vdots &\vdots&\vdots&\\
0&0&0&0&\dots&0&1&0\\
0&0&0&0&\dots& 1&0&1\\
0&0&0&0&\dots&0&1&1
\end{array} \right )}
\end{equation}
{\it The boundary conditions}

For $k=t$ we only use the loop $ABEDA$ in figure 3 to obtain the boundary equations
\begin{eqnarray}
I_{t-1}(i) =(2h+1)I_t(i)- hI_t(i-1)-hI_t(i+1)\qquad \hbox{ for $i<m$}\\
I_{t-1}(m) = (h+1)I_t(m)-hI_t(m-1) \qquad \qquad \qquad \quad
\end{eqnarray}
or in matrix form
\begin{eqnarray}
I_{t-1}=[(2h+1)U_m -&hV_m]I_t
\end{eqnarray}
with a similar equation for $k=-s$.

\subsection{3.2 The recurrence relation}

$V_m$ has eigenvalues $v_i = \cos(2\theta_i)$ and eigenvectors $\psi_i, i=1,2,\dots,m$. The $j^{th}$ component of $\psi_i$ is given by [7]
\begin{equation}
\psi_i(j) =  \sin[\frac{(2i-1)\pi j}{2m+1}] =\sin(2j\theta_i) .
\end{equation}

 Let $\Psi$ be the matrix with $i^{th}$ row $\psi_i$ and define $X_k=\Psi I_k$.  Inverting this relation and using (16)
\begin{equation}
R_{fan}= \frac{r_0}J\left(\sum_{i=1}^m  X_q(i) S_i(y_q) - \sum_{i=1}^m  X_{-p}(i) S_i(y_p) \right) ,
\end{equation}
where
\begin{equation}
S_i(y) = \sum_{j=y+1}^m(\Psi^{-1})_{ji} = \frac 4 {2m+1} \sum_{j=y+1}^m \sin(2j\theta_i) = \frac 2 {2m+1}\frac {\cos[(2y+1)\theta_i]}{\sin \theta_i} .
\end{equation}

Multiplying (21) on the left by $\Psi$, noting that $\Psi V_m$ is diagonal with diagonal elememts $v_i$, and taking the $i^{th}$ component
\begin{equation}
X_{k+1}(i) = u_i X_k(i) -X_{k-1}(i) -hJ\delta_{k,x}\zeta_i(y) ,
\end{equation}
where $u_i = 2h+2 -h v_i$ and
\begin{equation}
\zeta_i(y)\equiv \psi_i (y+1)-\psi_i(y)= 2 \sin\theta_i\cos[(2y+1)\theta_i] .
\end{equation}
 Applying $\Psi$ to (25) and taking the $i^{th}$ component
\begin{equation}
 X_{t-1}(i) = (u_i-1)X_t(i) \quad \hbox{and similarly} \quad  X_{-s+1}(i) = (u_i-1)X_{-s}(i) .
\end{equation}

\subsection{3.3 Solving the recurrence relation}

For $k\ne x$ the general solution of (29) is a linear combination of $\lambda_i^k$ and $\bar \lambda_i^k$ where
where $\lambda_i$ and $\bar \lambda_i$ are solutions of (1) in terms of which $\lambda_i +\bar \lambda_i = u_i$ and $\lambda_i \bar\lambda_i=1$.
The  coefficients depend on the region.
\begin{eqnarray}
X_k(i) &= A_i \lambda_i^k + \bar A_i \bar\lambda_i ^k \qquad \hbox{for} \qquad -p\le k \le q\\
X_k(i) &= B_i \lambda_i^k + \bar B_i \bar\lambda_i ^k \quad \hbox{for} \qquad q\le k \le t  \\
X_k(i) &= C_i \lambda_i^k + \bar C_i \bar\lambda_i ^k \qquad \hbox{for} \qquad -s\le k \le -p
\end{eqnarray}
Matching the solutions at $k=q$ and $k=-p$
\begin{eqnarray}
(A_i-B_i)\lambda_i^{2q} +\bar A_i- \bar B_i&=0\qquad \hbox{and} \qquad (\bar A_i -\bar C_i)\lambda_i^{2p} +A_i-C_i=0 .
\end{eqnarray}
Subsituting in the boundary equations (31)
\begin{equation}
\bar B_i =B_i\lambda_i^{2t+1} \qquad \hbox{and} \qquad C_i = \bar C_i \lambda_i^{2s+1} .
\end{equation}
The final two relations arise from the $k=q$ and $k=-p$ radial lines where the current $J$ is input and output. Using (29) with $x=k=q$ and secondly with $x=k=-p$, in the second case replacing $J$ by $-J$
\begin{equation}
\bar B_i- \bar A_i = -\frac{ hJ \lambda_i^q \zeta_i (y_q)}{\lambda_i- \bar\lambda_i} \qquad \hbox{and} \qquad \bar C_i -\bar A_i= -\frac{ hJ \bar\lambda_i^p \zeta_i (y_p)}{\lambda_i- \bar\lambda_i} .
\end{equation}
Solving equations (35), (36) and (37) for $A_i$ and $\bar A_i$ and substituting in (32) gives
\begin{equation}
X_q(i) = \frac {hJ [\alpha \zeta_i(y_q)-\beta\zeta_i(y_p)]}{(\lambda_i-\bar\lambda_i)D_i}\quad \hbox{and}\quad
X_{-p}(i) = -\frac {hJ [\gamma\zeta_i(y_p)-\beta\zeta_i(y_q)]}{(\lambda_i-\bar\lambda_i)D_i} ,
\end{equation}
where $D_i=\lambda_i^{ n} - \lambda_i^{-n}$.
\begin{eqnarray}\alpha &=(\lambda_i^{t-q+\frac{1}{2}} + \bar\lambda_i^{t-q+\frac{1}{2}})(\lambda_i^{s+q+\frac{1}{2}} + \bar\lambda_i^{s+q+\frac{1}{2}})\\
\beta&= (\lambda_i^{t-q+\frac{1}{2}} + \bar \lambda_i^{t-q+\frac{1}{2}})(\lambda_i^{s-p+\frac{1}{2}}+\bar\lambda_i^{s-p+\frac{1}{2}})\\
\gamma&= (\lambda_i^{t+p+\frac{1}{2}}+\bar\lambda_i^{t+p+\frac{1}{2}})(\lambda_i^{s-p+\frac{1}{2}}+\bar\lambda_i^{s-p+\frac{1}{2}})
\end{eqnarray}

Substituting (38) into (27) gives the required result (2) .

\section{4. Derivation of the cobweb formula (6)}
All that needs to be done to obtain the cobweb formula is to replace equations (36)  by
\begin{eqnarray}
B_i \lambda_i^t +\bar B_i \bar\lambda_i ^t = C_i \lambda_i^{-s-1} +\bar C_i\lambda_i^s , \\
B_i\lambda_i^t +\bar B_i \bar\lambda_i ^t = C_i\lambda_i^{-s} +\bar C_i\lambda_i^{s+1} .
\end{eqnarray}
Solving equations (35), (37), (42) and (43) for $A_i$ and $\bar A_i$ and
substituting in (32) gives (38)  where now, using $s+t+1=n$, $D_i=\lambda_i^{\frac{1}{2}n}- \lambda_i^{-\frac{1}{2}n}$ and
\begin{equation}
\alpha =\gamma=\lambda_i^{\frac{1}{2}n} +\lambda_i^{-\frac{1}{2}n}\qquad \hbox{and} \qquad  \beta=\lambda_i^{\frac{1}{2}n -p-q} +\lambda_i^{-\frac{1}{2}n+p+q } .
\end{equation}
Substituting (38), with these values of $\alpha,\beta,\gamma$ and $D_i$, into (27) gives the required result (6)
\begin{figure*}
\begin{center}
\includegraphics[width=9cm,bb=0 0 508 432]{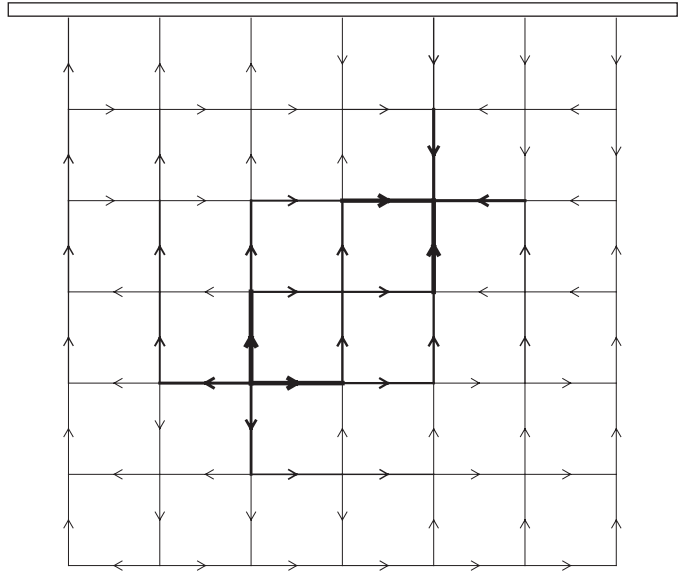}
\caption{ \it  The flow on a $6\times 7$ fan network with input node at $(-1, 2)$ and output node at $(1,4)$. The bar above the network has zero resistance. The line width represents the current magnitude.}
\end{center}
\end{figure*}

\section{5. Summary and discussion.}
The method of Tan, Zhou and Yang [10] has been used to derive a formula (2) for the resistance of a fan network  (see fig.1). The formula is for the resistance between two nodes in general position. By changing only the boundary condition (36) we also reproduce the result (6) of Izmailian, Kenna and Wu [7] for the related cobweb network. Their method is quite different from ours in that it requires the diagonalisation of two matrices instead of one. Also significant further work is required to reduce their formula from two to a single summation.  A further advantage of the present method is that the current distribution is also determined (see for example figure 4)..
Thus
\begin{equation}
I_k(i) = \sum_{j=1}^m(\Psi^{-1})_{ij}X_k(j) =\frac{4}{2m+1}\sum_{j=1}^m \sin(2j\theta_i)X_k(j) ,
\end{equation}
where $X_k(j)$ is given by (32),(33) and (34). The coefficients are determined at the same time as solving for $A_i$ and $\bar A_i$.

When both input and output nodes are on the same radial line and $s=t$ it is found that the fan and cobweb networks have the same resistance. The two networks are related by connecting corresponding nodes in the radial boundaries. It is argued that because of the symmetry there will be no flow in the connecting bonds  and hence the resistance equality.

The method of Tan et al [10] used here splits the derivation into three parts. The first creates a matrix relation between the current distributions on three successive radial lines. The second part diagonalises the matrix relation to produce a recurrence relation involving only variables on the same transverse line.  The relation is second order with constant coefficients and is solved in the third part. Basically the method reduces the problem from two dimensions to one dimension.

Equation (20) can be written in the form
\begin{equation}
(\Delta_k^2 +h\Delta_i^2)I_k(i) =-hJ \delta_{k,x} \epsilon_{i,y}
\end{equation}
which when $k\ne x$ reduces to the discrete Laplace equation. The method described here is therefore applicable to any problem involving the Laplacian.

\section*{Acknowledgment}
This work is supported by Jiangsu Province Education Science Plan Project (No. D/2013/01/048), the Research Project for Higher Education Research of Nantong University (No. 2012GJ003).

\end{document}